\documentclass[aps,reprint,groupedaddress]{revtex4-2}

\usepackage{graphicx}
\usepackage{color}
\usepackage{dcolumn}
\usepackage{bm}
\usepackage{amsmath}
\usepackage{here}

\begin{document}

\title{Aspect-ratio-dependent void formation in active rhomboidal and elliptical particle systems}

\author{Motoya Suzaka}
\affiliation{Department of Physics, Chiba University, Chiba 263-8522, Japan}
\author{Hiroaki Ito}
\affiliation{Department of Physics, Chiba University, Chiba 263-8522, Japan}
\author{Hiroyuki Kitahata}
\email{kitahata@chiba-u.jp}
\affiliation{Department of Physics, Chiba University, Chiba 263-8522, Japan}
\date{\today}

\begin{abstract}
We execute a numerical simulation on active nematics with particles interacting by an excluded volume effect.
The systems with rhomboidal particles and that with elliptical particles are considered in order to investigate the effect of the direct contact of particles.
In our simulation, the void regions, where the local number density is almost zero, appear in both systems when the aspect ratio of the particles is high.
We focused on the relationship between the void regions and the particle orientation of the bulk.
The particle number density, particle orientation, topological defects, and void regions are analyzed for different aspect ratios in both systems.
The systems with rhomboidal particles have characteristic void sizes, which increase with an increase in the aspect ratio.
In contrast, the distribution of the void-region size in the systems with elliptical particles is broad.
The present results suggest that the void size in the systems with rhomboidal particles is determined by the correlation length of the particle orientational field around the void regions, while that might be determined by the system size in the systems with elliptical particles.
\end{abstract}

\maketitle

\section{Introduction}
Active matter has attracted interest since a group of self-propelled particles, which can transform chemical energy into kinetic energy, exhibits characteristic ordering in the greater spatio-temporal scales than the characteristic scale of comprising particles\cite{SK001,SK002,SK003}.
Active matter can be classified into several types with respect to the symmetry of the particles; scalar, polar, nematic, and chiral active matter\cite{SK010}.
Here, we focus on the active nematic systems, in which particles interact with each other based on dyad rotational symmetry.
They have been studied in both experiments and numerical simulations\cite{SK016}.
Microtubules driven by motor proteins on substrates\cite{SK051,SK031,SK059,SK060}, colonies of rod-like bacteria\cite{SK017,SK061,SK062,SK053,SK054}, and elongated cells\cite{SK055,SK018,SK052,SK056} are the typical experimental systems for active nematics.
In such systems, locally aligned structures of particles and inhomogeneous density distribution often emerge.
Characteristic patterns in coarse-grained orientational fields appear as vortices or bent bands with topological defects, i.e., singular points\cite{SK020,SK031,SK033}.
It was reported that motion of topological defects due to particle activity breaks the long-range nematic order and leads spatio-temporal dynamics of the orientation patterns\cite{SK034,SK035,SK036}.
Emergence of characteristic density patterns has been reported;
In low-density systems, clustering and phase separation appear\cite{SK025, SK026,SK027,SK063}, and in high-density systems, topological defects induce the density inhomogeneity through the inflow or outflow of particles around them\cite{SK044,SK062,SK046,SK047,SK048}.
In a middle-density system, the void regions are observed, where local particle number density is much lower than the mean number density of the whole system.
Shi and Ma simulated an active nematics system with elliptical particles at a middle density and claimed that void regions are generated through the collision of two $+1/2$ topological defects\cite{SK049}.

In this study, we focus on the middle-density system and investigate the size of void regions, which should be essential to determine the properties of the whole system.
We are especially interested in whether the size of the void regions is determined by the system size or by other properties of particles and bulk.
The effect of the aspect ratio has been conventionally focused on since the aspect ratio should be the leading term in the low-density models with hydrodynamic interaction.
However, in our middle-density system, the particle shape should be another important factor to determine the size of void regions because particles interact by direct collisions.
Actually, it was reported that the pattern dynamics of active nematics with an excluded volume effect are changed by the particle geometry in active nematics.
The transition in the collective behavior was reported by changing the particle aspect ratio in the multi-particle systems with elliptical particles\cite{rodsAranson} and with rod-like particles\cite{rodsWensink}.
Moran et al. investigated the effect of the particle anisotropy on the clustering dynamics by comparing the circular and regular polygonal particle systems\cite{SK015_5,SK015_6}.
We speculate that details in geometry, as well as aspect ratio, may play an essential role for emergence of the spatio-temporal patterns in active nematic systems. 
Therefore, we adopt the two different particle shapes, rhomboidal and elliptical particles, in order to investigate the effect of the difference in direct contact among particles.
We construct an active nematics model based on the one by Shi and Ma\cite{SK049} and perform numerical simulation.
The aspect ratio of particles is changed as a parameter to check the effect of the magnitude of the nematic interaction.
The local nematic order, the area of void regions, and the winding numbers of topological defects are analyzed.
Through these analyses, the possible relation between the characteristic void size and the orientational structure of the void regions is discussed with geometrical consideration.

The paper consists of the sections as follows.
In Sec.\,\ref{SecTwo}, we explain the methods of the simulation and the data analyses.
In Sec.\,\ref{SecThree}, we show the results of the numerical simulation, focusing on the particle-shape dependence in the local number density, the local orientation, the topological defects, and the void regions.
In Sec.\,\ref{SecFour}, we discuss the mechanism to determine the characteristic size of void regions, considering the orientational structure around the void regions.
In Sec.\,\ref{SecFive}, We summarize our study.

\section{Method}
\label{SecTwo}

\subsection{Simulation model}
\label{SecTwoA}

We consider a two-dimensional squared system with a periodic boundary condition, where the length of the sides is $L^{(\mathrm{sys})}$ and the area is thus $(L^{(\mathrm{sys})})^2$.
The system includes $N$ self-propelled nematic particles with the same shapes.
As the shape of the nematic particles, two cases, i.e., rhomboid or ellipse, are considered.
We set $\boldsymbol{x}_{i,t}$ to be the coordinates of the center of mass (COM) of the $i$-th ($i=1, \dots ,N$) particle at a discrete time step $t$, and $\theta_{i,t}$ to be the angle of the traveling direction, which meets one of the two long-axis directions of the particle.
We also introduce the variable $\Lambda_{i,t}$, which represents the memory on the overlap with other particles at the last step.

The dynamics of the particles is described as follows.
The particles interact with each other only by the excluded volume effect.
The trial motion is described with a tentative position $\boldsymbol{x}'_{i,t}$ and angle $\theta'_{i,t}$ as
\begin{align}
\label{centermass_trial}
\boldsymbol{x}'_{i,t}=&\boldsymbol{x}_{i,t}+D^{(t)}(1+\eta^{(t)}_{i,t})\boldsymbol{e} \left( \theta_{i,t} \right)\nonumber\\
&+\Lambda_{i,t}D^{(r)}(1+\eta^{(r)}_{i,t})\boldsymbol{e} \left( \pi\eta^{(\varphi)}_{i,t} \right) ,
\end{align}
\begin{equation}
\label{theta_trial}
\theta'_{i,t}=\theta_{i,t}+\pi\zeta_{i,t}+\Lambda_{i,t}\pi D^{(\theta)}\eta^{(\theta)}_{i,t},
\end{equation}
and then the overlap with other particles is checked.
The trial motion is rejected if any overlap is detected, while it is accepted if overlap is not detected.
In other words, the position, the angle, and the memory at the next time step are determined as 
\begin{align}
\label{centermass_actual}
(\boldsymbol{x}_{i,t+1},\ & \theta_{i,t+1},\ \Lambda_{i,t+1})\nonumber\\
=&\left\{
\begin{array}{ll}
(\boldsymbol{x}_{i,t},\ \theta_{i,t},\ 1),&\text{if overlaps are detected at the trial,}\\
(\boldsymbol{x}' _{i,t},\ \theta'_{i,t},\ 0),&\text{otherwise.}\\
\end{array}
\right.
\end{align}
$\eta^{(t)}_{i,t}$, $\eta^{(r)}_{i,t}$, $\eta^{(\varphi)}_{i,t}$ and $\eta^{(\theta)}_{i,t}$ in Eqs.~(\ref{centermass_trial}) and (\ref{theta_trial}) are stochastic variables that obey the uniform distribution in $[-1,1]$.
$2D^{(t)}$ corresponds to the maximum distance of the motion in the traveling direction of the particle.
$2D^{(r)}$ and $\pi D^{(\theta)}$ correspond to the maximum distance of the motion in a random direction and the maximum angle of a random rotation, respectively, in the case that the overlap is detected at the last time step.
$\boldsymbol{e}(\theta)=(\cos{\theta},\sin{\theta})$ is a unit vector in the direction of $\theta$.
$\zeta_{i,t}$ is a stochastic variable representing the switching of the traveling direction with a characteristic time $\tau$,
\begin{align}
\label{zeta_actual}
\zeta_{i,t}=\left\{
\begin{array}{ll}
1,&\text{probability of }1/\tau,\\
0,&\text{probability of }1-1/\tau.\\
\end{array}
\right.
\end{align}

In this study, the aspect ratio of the particle $a$ is varied from $1.5$ to $6.0$ as a parameter under the constraint that the particle area $s$ is constant and common for rhomboidal and elliptical particles.
That is to say, the long-axis lengths $l^{(\mathrm{r})}$ and $l^{(\mathrm{e})}$ for rhomboidal and elliptical particles are determined by $s=(l^{(\mathrm{r})})^2/(2a)$, and $s=\pi (l^{(\mathrm{e})})^2/(4a)$.
The area of the particles is fixed to be $s=\pi/23$.
The other parameters are fixed as $N=4140$, $L^{(\mathrm{sys})}=30$, $D^{(t)}=0.006$, $D^{(r)}=0.002$, $D^{(\theta)}=0.015$, and $\tau=6$.
The time evolution processes are repeated up to $t=2\times10^7$.
Ten times numerical simulations are performed for each parameter set, with different initial conditions and different stochastic noise series.
Statistic quantities are calculated from the data every $10^4$ time step for $t \geq 1 \times 10^7$ in the ten numerical results.

\subsection{Data Analysis}
\label{SecTwoB}
To measure the local number density, the local nematic order, and the local mean orientation, $M_{x} \times M_{y}$ measurement points are set on a regular grid, where $M_{x}$ and $M_{y}$ are even numbers and we set $M_{x}=M_{y}=64$ in the present analysis.
The measurement points are indexed as $\boldsymbol{j}=(j_{x},j_{y})$ ($j_{x}=1, \dots ,M_{x}$; $j_{y}=1, \dots ,M_{y}$), and the corresponding positional vectors are $\boldsymbol{y}_{\boldsymbol{j}}=(L^{\mathrm{(sys)}}j_{x}/M_{x},L^{\mathrm{(sys)}}j_{y}/M_{y})$.
Under a given particle configuration $\{ \boldsymbol{x}_{i,t} \}$, local number density $\rho_{\boldsymbol{j},t}$ at the measurement point $\boldsymbol{j}$ at the time step $t$ is defined as
\begin{align}
\label{local_numberdensity}
\rho_{\boldsymbol{j},t}&=\frac{ \int \sum_{i=1}^{N} \delta(\boldsymbol{r}-\boldsymbol{x}_{i,t}) W(\boldsymbol{r}-\boldsymbol{y}_{\boldsymbol{j}})d\boldsymbol{r} }{ \int W(\boldsymbol{r}-\boldsymbol{y}_{\boldsymbol{j}}) d\boldsymbol{r} }\nonumber \\
&\simeq\frac{\sum_{i=1}^{N} W(\boldsymbol{x}_{i,t}-\boldsymbol{y}_{\boldsymbol{j}}) }{ \int W(\boldsymbol{r}-\boldsymbol{y}_{\boldsymbol{j}}) d\boldsymbol{r} }.
\end{align}
Here, $W(\boldsymbol{r})$ is a kernel function
\begin{align}
\label{weighting}
W(\boldsymbol{r})=\left\{
\begin{array}{ll}
0,& \left| \boldsymbol{r} \right| >1,\\
1-\left| \boldsymbol{r} \right| ,& \left| \boldsymbol{r} \right| \leq 1,
\end{array}
\right.
\end{align}
and $\delta(\cdot)$ represents the Dirac's delta function.

The local nematic order $P_{\boldsymbol{j},t}$ at the measurement point $\boldsymbol{j}$ at the time step $t$ is defined as
\begin{equation}
\label{local_nemop}
P_{\boldsymbol{j},t}=\left| \det Q_{\boldsymbol{j},t} \right| .
\end{equation}
Here, $Q_{\boldsymbol{j},t}$ is defined as
\begin{equation}
\label{local_directionmatrix}
Q_{\boldsymbol{j},t}=\left(\begin{array}{cc} H^{(\cos)}_{\boldsymbol{j},t} & H^{(\sin)}_{\boldsymbol{j},t} \\ H^{(\sin)}_{\boldsymbol{j},t} & -H^{(\cos)}_{\boldsymbol{j},t} \\ \end{array} \right),
\end{equation}
where
\begin{align}
\label{local_cos2theta}
H^{(\cos)}_{\boldsymbol{j},t}&=\frac{ \int \sum_{i=1}^{N} \delta(\boldsymbol{r}-\boldsymbol{x}_{i,t}) W(\boldsymbol{r}-\boldsymbol{y}_{\boldsymbol{j}})\cos{2\theta_{i,t}}\, d\boldsymbol{r} }{ \int W(\boldsymbol{r}-\boldsymbol{y}_{\boldsymbol{j}})\, d\boldsymbol{r} } \nonumber \\
&\simeq\frac{\sum_{i=1}^{N} W(\boldsymbol{x}_{i,t}-\boldsymbol{y}_{\boldsymbol{j}})\cos{2\theta_{i,t}}\,}{ \int W(\boldsymbol{r}-\boldsymbol{y}_{\boldsymbol{j}})\, d\boldsymbol{r} },
\end{align}
\begin{align}
\label{local_sin2theta}
H^{(\sin)}_{\boldsymbol{j},t}&=\frac{ \int \sum_{i=1}^{N} \delta(\boldsymbol{r}-\boldsymbol{x}_{i,t}) W(\boldsymbol{r}-\boldsymbol{y}_{\boldsymbol{j}})\sin{2\theta_{i,t}}\, d\boldsymbol{r} }{ \int W(\boldsymbol{r}-\boldsymbol{y}_{\boldsymbol{j}})\, d\boldsymbol{r} }\nonumber \\
&\simeq\frac{\sum_{i=1}^{N} W(\boldsymbol{x}_{i,t}-\boldsymbol{y}_{\boldsymbol{j}})\sin{2\theta_{i,t}} }{ \int W(\boldsymbol{r}-\boldsymbol{y}_{\boldsymbol{j}})\, d\boldsymbol{r} }.
\end{align}
The local mean orientation $\phi_{\boldsymbol{j},t}$ at the measurement point $\boldsymbol{j}$ at the time step $t$ is defined as
\begin{align}
\label{local_orientation direction}
H_{0}e^{2\sqrt{-1}\phi_{\boldsymbol{j},t}}=H^{(\cos)}_{\boldsymbol{j},t}+\sqrt{-1}H^{(\sin)}_{\boldsymbol{j},t},\quad (H_0>0).
\end{align}
It should be noted that $\phi_{\boldsymbol{j},t}$ cannot be defined at the measurement points with $H^{(\cos)}_{\boldsymbol{j},t}=H^{(\sin)}_{\boldsymbol{j},t}=0$.

Next, to characterize topological defects and void regions, $(M_{x}/2) \times (M_{y}/2)$ coarse measurement points are set on a regular grid.
The coarse measurement points are indexed as $\boldsymbol{k}=(k_{x},k_{y})$ ($k_{x}=1, \dots ,(M_{x}/2); k_{y}=1, \dots ,(M_{y}/2)$), and the corresponding positional vectors are $\boldsymbol{z}_{\boldsymbol{k}}=(2L^{\mathrm{(sys)}}k_{x}/M_{x},2L^{\mathrm{(sys)}}k_{y}/M_{y})$.
It should be noted that $\boldsymbol{y}_{2\boldsymbol{k}}=\boldsymbol{z}_{\boldsymbol{k}}$ holds.

A coarse measurement point indexed by $\boldsymbol{k}$ is called as a ``void point'' if the minimum of the nine values $\rho_{\boldsymbol{j}, t}$ for $j_x= 2k_x, 2k_x \pm 1$ and $j_y = 2k_y, 2k_y \pm 1$ is less than $0.3 \rho^{(\mathrm{sys})}$, where $\rho^{(\mathrm{sys})}=N/(L^{\mathrm{(sys)}})^{2}$.
A ``void region'' is defined as the eight-neighboring connected ``void points''.
The number of void regions at the time step $t$ is represented as $\mathcal{N}_{t}^{(\mathrm{void})}$.
The area of the $m$-th void region at the time step $t$ is defined as $S_{m,t}^{(\mathrm{void})}=\chi_{m,t} (L^{(\mathrm{sys})})^2/(M_{x}M_{y}/4)$ ($m=1,\dots,\mathcal{N}_{t}^{(\mathrm{void})}$), where $\chi_{m,t}$ is the number of void points belonging to the $m$-th void region. 
The mean area of void regions weighted by the void region area at the time step $t$ is defined as
\begin{equation}
\label{voidArea}
\bar{S}^{(\mathrm{void})}_{t}=\frac{ \sum_{m} (S^{(\mathrm{void})}_{m,t})^2 }{ \sum_{m} S^{(\mathrm{void})}_{m,t} }.
\end{equation}
Then, $\left< \bar{S}^{(\mathrm{void})} \right>$ is defined as the mean of $\bar{S}^{(\mathrm{void})}_{t}$ with respect to $t$ and the ten trials.

The winding number $\Phi_{\boldsymbol{k},t}$ at the coarse measurement point indexed by $\boldsymbol{k}$ is calculated as follows.
First, the indices of the surrounding eight measurement points around the focused coarse measurement point are set as $\boldsymbol{j} = \boldsymbol{J}_{1}^{(\boldsymbol{k})}, \dots , \boldsymbol{J}_{8}^{(\boldsymbol{k})}$, so that they surround the focused point counterclockwise.
The winding number of the void region indexed by $m$ at the time step $t$ is calculated almost in the same manner except for the choice of the path.
The indices of the measurement points surrounding the focused void region are set as $\boldsymbol{j}_{t} = \boldsymbol{J}_{1,t}^{(m)}, \dots , \boldsymbol{J}_{g,t}^{(m)}$($g=\nu_{m,t}$), so that they surround the focused void region indexed with $m$ counterclockwise.
Here, $\nu_{m,t}$ is the number of the surrounding measurement points around the $m$-th void region.
We define the local orientation difference between the neighboring coarse measurement points $\boldsymbol{J}'$ and $\boldsymbol{J}''$ as
\begin{align}
\label{windingNumber}
\Delta\phi_{\boldsymbol{J}',\boldsymbol{J}'',t} = \phi_{\boldsymbol{J}',t}-\phi_{\boldsymbol{J}'',t},\quad(-\frac{\pi}{2}<\Delta\phi_{\boldsymbol{J}',\boldsymbol{J}'',t}<\frac{\pi}{2}).
\end{align}
Then, the winding number is calculated as 
\begin{align}
\label{windingNumber_2}
\Phi_{\boldsymbol{k},t} = \frac{1}{2\pi} \sum_{\ell = 1}^{8} \Delta\phi_{\boldsymbol{J}_{\ell}^{(\boldsymbol{k})},\boldsymbol{J}_{\ell -1}^{(\boldsymbol{k})}},
\end{align}
for coarse measurement points, and that of the void regions is also calculated as 
\begin{align}
\label{windingNumber_V2}
\Psi_{m,t} = \frac{1}{2\pi} \sum_{\ell = 1}^{\nu_{m,t}} \Delta\phi_{\boldsymbol{J}_{\ell,t}^{(m)},\boldsymbol{J}_{\ell -1,t}^{(m)}},
\end{align}
for void regions.
Here, it should be noted that we set $\boldsymbol{J}_{0}^{(\boldsymbol{k})} = \boldsymbol{J}_{h}^{(\boldsymbol{k})}$ and $\boldsymbol{J}_{0,t}^{(m)} = \boldsymbol{J}_{g,t}^{(m)}$, and islands of the coarse measurement points with $\rho_{\boldsymbol{j},t} \geq 0.3 \rho^{(\mathrm{sys})}$ inside of void regions are ignored and only the coarse measurement points along the peripheries of the void regions are considered.
\section{Results}
\label{SecThree}
\begin{figure*}
\includegraphics{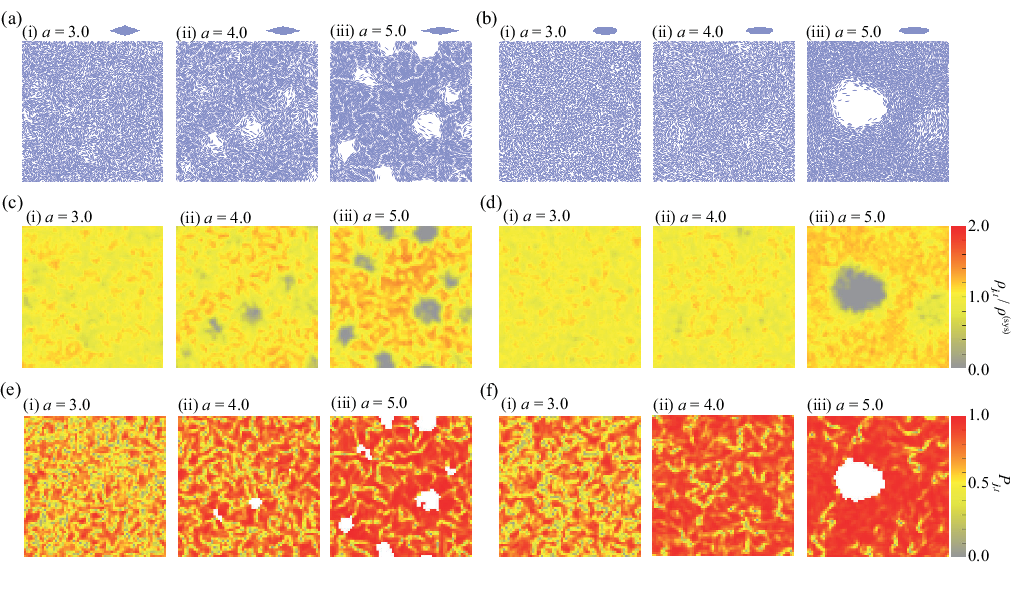}
\caption{\label{snapshots}
Representative results of the numerical simulations for the systems with (a,c,e) rhomboidal and (b,d,f) elliptical particles at $t=1.9\times 10^7$. Aspect ratio is (i) $a=3.0$, (ii) $a=4.0$, and (iii) $a=5.0$.
(a,b) Snapshots. The particle shape is illustrated on the upper side. (c,d) Local number density $\rho_{\boldsymbol{j},t} / \rho^{(\mathrm{sys})}$. (e,f) Local nematic order $P_{\boldsymbol{j},t}$, where white areas indicate the void regions in (e,f).
}
\end{figure*}
Figure~\ref{snapshots} shows the results of the numerical simulation.
Snapshots at $t=1.9\times10^7$ for the systems with rhomboidal and elliptical particles, with the aspect ratios $a=3.0$, $4.0$, and $5.0$ are shown in Figs.~\ref{snapshots}(a,b).
Figures~\ref{snapshots}(c--f) show the corresponding local number density $\rho_{\boldsymbol{j},t} / \rho^{(\mathrm{sys})}$ (Figs.~\ref{snapshots}(c,d)), and the local nematic order $P_{\boldsymbol{j},t}$ (Figs.~\ref{snapshots}(e,f)).
Both in the systems with rhomboidal and elliptical particles, the void regions that are shown as the white regions in Figs.~\ref{snapshots}(e,f) are observed for the higher aspect ratio.
In addition, the particles are more aligned for the higher aspect ratio as shown in Figs.~\ref{snapshots}(e,f).
The void regions are observed for rhomboidal particles with $a=4.0$ and $5.0$ and elliptical particles with $a=5.0$.
Multiple small void regions are observed in the systems with rhomboidal particles, while a single large void region is observed in the systems with elliptical particles.
\begin{figure}
\includegraphics{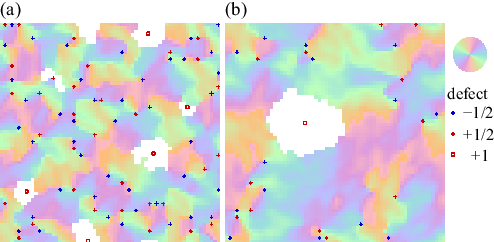}
\caption{\label{defectMap}
The local mean orientations $\phi_{j,t}$ with the positions of the topological defects in the systems with (a) rhomboidal and (b) elliptical particles for $a=5.0$ at $t=1.9\times 10^7$.
White areas indicate the void regions.
}
\end{figure}
Figures~\ref{defectMap}(a,b) show the local mean orientations $\phi_{j,t}$ with the positions of the topological defects in the systems with rhomboidal (Fig.~\ref{defectMap}(a)) and elliptical (Fig.~\ref{defectMap}(b)) particles for $a=5.0$.
In these representative results, most of topological defects appear with the winding number $\pm{1/2}$ and the void regions appear with the winding number $1$.

\begin{figure}
\includegraphics{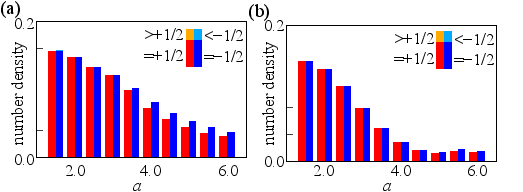}
\caption{\label{WR}
Number density of the topological defects depending on the aspect ratio $a$ for the systems with (a) rhomboidal and (b) elliptical particles.
The colors indicate the winding numbers.
The fractions of the defects with winding numbers more than $+1/2$ or less than $-1/2$ are too small to recognize.
}
\end{figure}
Figure~\ref{WR} shows the number density of defects depending on the aspect ratio $a$ for both the systems with rhomboidal and elliptical particles.
It decreases with an increase in $a$, and the systems with rhomboidal particles include more defects than those with elliptical particles for the same aspect ratio.
Almost all topological defects have the winding number $+1/2$ or $-1/2$, and the defects with the winding numbers more than $+1/2$ or less than $-1/2$ are rarely observed.
\begin{figure}
\includegraphics{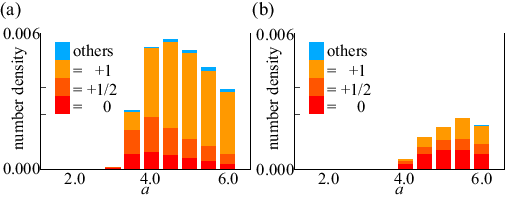}
\caption{\label{VWR}
Number density of the void regions depending on the aspect ratio $a$ for the systems with (a) rhomboidal and (b) elliptical particles.
The colors indicate winding numbers.
}
\end{figure}
Figure~\ref{VWR} shows the number density of void regions with various winding numbers for both in the systems with rhomboidal and elliptical particles.
Void regions are not observed for $a \leq 3$ in the systems with rhomboidal particles and for $a \leq 4$ in those with elliptical particles.
Nonmonotonical changes of the number density of void regions with an increase in $a$ are observed. 
The number density of void regions is higher in the systems with rhomboidal particles than that in the systems with elliptical particles.
Almost all void regions have the positive or zero winding numbers.
The fraction of the winding number $+1$ increases with an increase in $a$.
The void regions with the winding numbers over $+1$ are rarely observed and those with the negative winding number are not observed.

\begin{figure}
\includegraphics{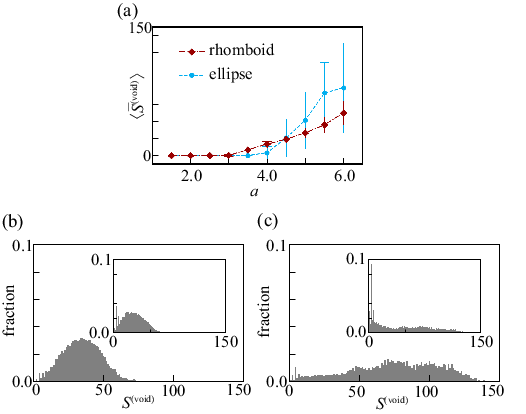}
\caption{\label{VoidS}
(a) Ensemble and time average of the void area $\left< \bar{S}^{(\mathrm{void})} \right>$.
Error bars indicate standard deviation.
Histogram of the void area $S^{(\mathrm{void})}$ for the systems with (b) rhomboidal and (c) elliptical particles.
Aspect ratio is $a=5.0$.
The fraction is weighted by the void region $S^{(\mathrm{void})}$.
The fraction with respect to the count is shown in the inset.
}
\end{figure}
Figure~\ref{VoidS}(a) shows the mean size of void regions $\left< \bar{S}^{\mathrm{(void)}} \right>$ against $a$ in both the systems with rhomboidal and elliptical particles.
$\left< \bar{S}^{\mathrm{(void)}} \right>$ increases with an increase in $a$ for the higher aspect ratio for both the systems.
In addition, for these $a$, the standard deviation of $\bar{S}^{\mathrm{(void)}}$ in the systems with rhomboidal particles is smaller than that in the systems with elliptical particles. 
The distribution of the void region area for both the systems with $a=5.0$ is shown in Figs.~\ref{VoidS}(b,c).
The distribution is weighted by the void region area $S^{(\mathrm{void})}$.
The distribution is narrower in the systems with rhomboidal particles than that in the systems with elliptical particles.

\section{Discussion}
\label{SecFour}
\begin{figure}
\includegraphics{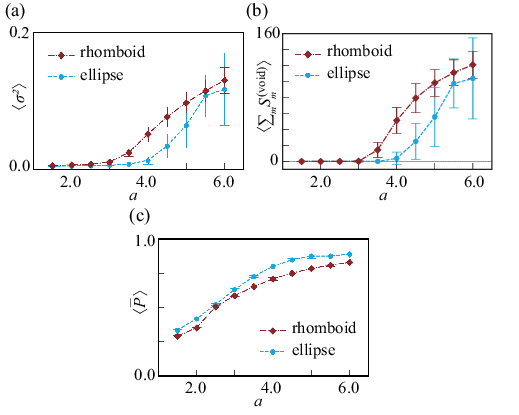}
\caption{\label{dependanceOfAspect}
Ensemble and time average of  (a) the local number density variance $\left< \sigma^{2} \right>$, (b) the total area of void regions $\left< \sum_{m} S^{(\mathrm{void})}_{m} \right>$, and (c) the local nematic order $\left< \bar{P} \right>$.
Error bars indicate standard deviation.
}
\end{figure}

Here, we focus on the emergence of the void regions depending on the particle shape and the aspect ratio.
To discuss the character of the void regions quantitatively, we define the variance $\sigma^{2}_{t}$ of the local number density $\rho_{\boldsymbol{j},t}$ with respect to $\boldsymbol{j}$ at the time step $t$ as
\begin{align}
\label{VarianceOfRho}
\sigma^{2}_{t}=\frac{\sum_{\boldsymbol{j}} (\rho_{\boldsymbol{j},t}-\bar{\rho}_{t})^2}{M_{x}M_{y}},
\end{align}
where $\bar{\rho}_{t}$ is the mean of $\rho_{\boldsymbol{j},t}$ with respect to $\boldsymbol{j}$ at the time step $t$.
$\left< \sigma^{2} \right>$ is defined as the mean of $\sigma^{2}_{t}$ with respect to $t$ and the ten trials.
Figure~\ref{dependanceOfAspect}(a) shows that the local density variance $\left< \sigma^{2} \right>$ changes from $\left< \sigma^{2} \right> \sim 0$ to finite values at $a=3.0$ and $4.0$ in the systems with rhomboidal and elliptical particles, respectively.
The transition in $\left< \sigma^{2} \right>$ corresponds to the emergence of the void regions as shown in Fig.~\ref{VWR}.
Figure~\ref{dependanceOfAspect}(b) shows the total area of void regions $\left< \sum_{m} S^{(\mathrm{void})}_{m} \right>$, which shows almost the same aspect ratio dependence as $\left< \sigma^{2} \right>$.
To discuss the relationship between the emergence of void regions and the particle orientational order, we define the mean local nematic order $\left< \bar{P} \right>$ as the mean of $\bar{P}_{t}$ with respect to $t$ and the ten trials, where $\bar{P}_{t}$ is the mean of $P_{\boldsymbol{j},t}$ with respect to $\boldsymbol{j}$ for the region with $\rho \geq 0.3\rho^{(\mathrm{sys})}$.
Figure~\ref{dependanceOfAspect}(c) shows that the local nematic order $\left< \bar{P} \right>$ gradually increases with an increase in $a$ both in the systems with rhomboidal and elliptical particles, even at the aspect ratio $3.0 \lesssim a \lesssim 4.0$, where $\left< \sigma^{2} \right>$ and $\left< \sum_{m} S^{(\mathrm{void})}_{m} \right>$ sharply increase.
This suggests that void regions can emerge without the drastic change in the orientation order.

The number density of void regions in the systems with elliptical particles is smaller than that in the systems with rhomboidal particles as shown in Fig.~\ref{VWR}.
We consider that the remarkable difference between the systems with rhomboidal particles and those with elliptical particles is the size and stability of the void regions.
In the systems with rhomboidal particles, the voids have a characteristic size, which is an increasing function of the aspect ratio, as indicated by Fig.~\ref{VoidS}.
Actually it shows that the distribution of $S^{(\mathrm{void})}$ with rhomboidal particles for $a=5.0$ is narrower than that with elliptical particles.
The total area of void regions is saturated around $a=5.0$ as shown in Fig.~\ref{dependanceOfAspect}(b).
The combination of the aspect-ratio-dependent characteristic size and the saturation of the total void area should lead the nonmonotonical dependence of the number density of void regions shown in Fig.~\ref{VWR}. 
In contrast, in the systems with elliptical particles, the void regions do not exist stably for the lower aspect ratio.
We consider that the density fluctuation becomes greater for the higher aspect ratio and the lower density regions due to the fluctuation are counted as the void regions.
This reflects that the fraction of the voids with zero winding number is higher for the systems with elliptical particles as shown in Fig.~\ref{VWR}(b).
For the higher aspect ratio $a \gtrsim 5.0$, the void size should be bounded by the system size.
This results in the saturated behavior of the number density of the void regions in the systems with elliptical particles.

\begin{figure}
\includegraphics{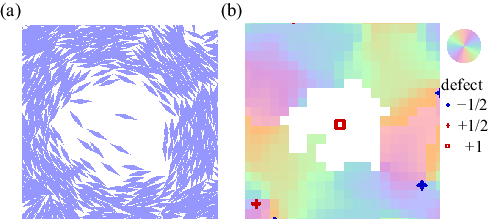}
\caption{\label{OneVoid}
Enlarged views around a void region with the winding number $+1$ for the systems with rhomboidal particles for $a=5.0$, which are taken from (a) Fig.~\ref{snapshots}(a)(iii) and (b) Fig.~\ref{defectMap}(a).
The area with $10\times 10$ is shown.
}
\end{figure}

\begin{figure}
\includegraphics{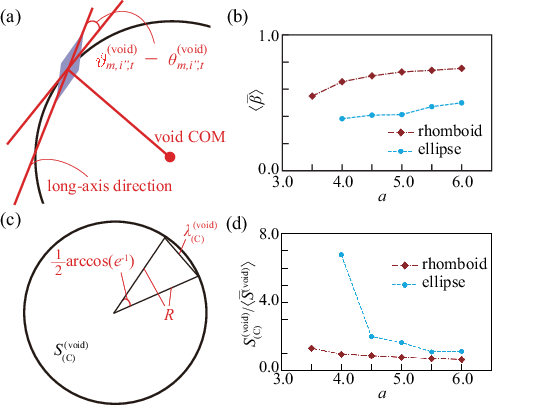}
\caption{\label{VoidSAC}
(a) Schematic illustration for the angle difference $\vartheta^{(\mathrm{void})}_{m, i'', t}-\theta_{m, i'' ,t}$.
(b) Ensemble and time average of the orientational correlation functions of particles and void peripheries $\left< \bar{\beta} \right>$.
(c) Schematic illustration for the estimation of void sizes $S^{\mathrm{(void)}}_{\mathrm{(C)}}$ from $\lambda^{(\mathrm{void})}_{\mathrm{(C)}}$.
The void region is considered as a circumscribed circle surrounded by edges whose length is given as the orientational correlation length.
The radius $R$ of the void region is calculated as in Eq. (\ref{VoidRadius}).
(d) Ratio $S^{\mathrm{(void)}}_{\mathrm{(C)}}/\left< \bar{S}^{(\mathrm{void})} \right>$ against $a$.
The data for $a<3.0$ in the systems with rhomboidal particles and $a<4.0$ in the systems with elliptical particles could not be evaluated because the number of voids was too small.
}
\end{figure}

Figure~\ref{VWR} shows that some void regions have the winding number $+1$.
Seeing particle orientation around the void regions, many void regions with winding number $+1$ are surrounded by particles whose long-axes are directed along with the peripheries of these void regions as shown in Fig.~\ref{OneVoid}.
We consider the size of the void regions may be associated with the particle orientation around them.
To discuss the characteristic void area, we evaluate the orientational correlation lengths.
\begin{figure}
\includegraphics{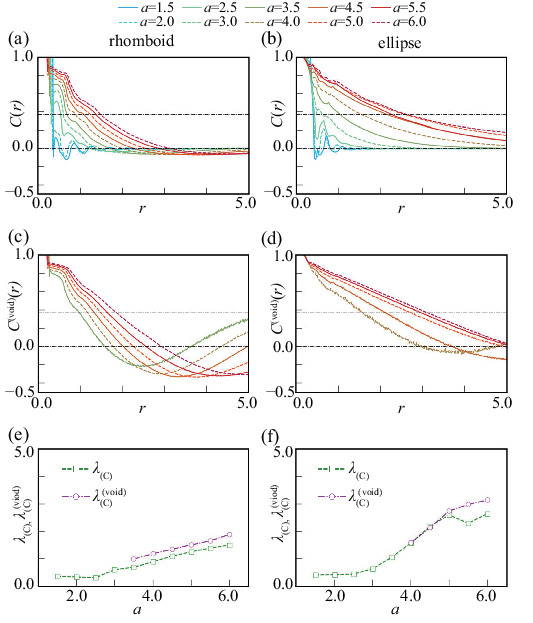}
\caption{\label{CL}
(a,b) Spatial correlation of the particle orientation $C(r)$ in the whole region for the systems with (a) rhomboidal and (b) elliptical particles.
(c,d) Spatial correlation of the particle orientation for particle pairs located around the same void regions $C^{\mathrm{(void)}}(r)$ for the systems with (c) rhomboidal and (d) elliptical particles.
(e,f) Correlation lengths in the whole region $\lambda_{(\mathrm{C})}$ and those for particle pairs located around the same void regions $\lambda^{(\mathrm{void})}_{(\mathrm{C})}$ depending on the aspect ratio $a$ for the systems with (e) rhomboidal and (f) elliptical particles.
}
\end{figure}
They are calculated from a spatial correlation function $C(r)$ of particle orientation which is defined as,
\begin{align}
\label{Correlation}
C(r) &= \left< \cos{2(\theta_{i,t}-\theta_{i',t})} \right>_{r<|\boldsymbol{x}_{i}-\boldsymbol{x}_{i'}|<r+\Delta r}.
\end{align}
Here, $\left< \cdot \right>_{r<|\boldsymbol{x}_{i}-\boldsymbol{x}_{i'}|<r+\Delta r}$ denotes the mean with respect to pairs of the $i$-th and $i'$-th ($i,i'=1, \dots ,N$) particles with distances $r<|\boldsymbol{x}_{i}-\boldsymbol{x}_{i'}|<r+\Delta r$ and with respect to $t$ and the ten trials.
The correlation length of the particle orientation $\lambda_{\mathrm{(C)}}$ is defined as the minimum length $r$ that satisfies $C(r) < e^{-1}$.
The spatial correlation of the particle orientation for the particle pairs located around the same void regions $C^{\mathrm{(void)}}(r)$ is also defined as
\begin{align}
\label{VoidCorrelation}
C^{\mathrm{(void)}}(r)&=\left< \cos{2(\theta_{i,t}-\theta_{i',t})} \right>_{ \left\{ r<|\boldsymbol{x}_{i}-\boldsymbol{x}_{i'}|<r+\Delta r \, |\, i,i' \in \boldsymbol{\mathcal{P}}_{m,t} \right\} }.
\end{align}
Here, $\boldsymbol{\mathcal{P}}_{m,t}$ is the set of the indices of the two particles whose distances from void points belonging to the $m$-th void region are both less than $1$ at the time step $t$.
$\left< \cdot \right>_{ \left\{ r<|\boldsymbol{x}_{i}-\boldsymbol{x}_{i'}|<r+\Delta r \, |\, i,i' \in \boldsymbol{\mathcal{P}}_{m,t} \right\}}$ denotes the mean of the values for the particle pairs, both of which are the elements of $\boldsymbol{\mathcal{P}}_{m,t}$ and whose distances are $r<|\boldsymbol{x}_{i}-\boldsymbol{x}_{i'}|<r+\Delta r$, and with respect to $t$ and the ten trials.
The correlation length $\lambda^{\mathrm{(void)}}_{(\mathrm{C})}$ of particle orientation is defined as the minimum length $r$ that satisfies $C^{\mathrm{(void)}}(r) < e^{-1}$.
Figure~\ref{CL} shows $C(r)$, $C^{\mathrm{(void)}}(r)$, $\lambda_{\mathrm{(C)}}$, and $\lambda^{\mathrm{(void)}}_{\mathrm{(C)}}$ for both the systems with rhomboidal and elliptical particles.
$\lambda_{\mathrm{(C)}}$ and $\lambda^{\mathrm{(void)}}_{\mathrm{(C)}}$ increase with an increase in $a$ in both the systems.
We also evaluate the orientational correlation between particles and void peripheries, which is defined as
\begin{align}
\label{EdgeCorrelation}
\beta_{m,t} &= \left< \cos{2(\vartheta^{\mathrm{(void)}}_{m,i'',t}-\theta_{i'',t})} \right>_{ \left\{ i'' \in \boldsymbol{\mathcal{E}}_{m,t} \right\} }.
\end{align}
Here, $\boldsymbol{\mathcal{E}}_{m,t}$ is the set of indices for the particles whose distances from void points belonging to the $m$-th void region are less than $1$ at the time step $t$, which indicates the particles at the void periphery.
$\left< \cdot \right>_{ \left\{ i'' \in \boldsymbol{\mathcal{E}}_{m,t} \right\} }$ denotes the mean of the values for the particles which are the elements of $\boldsymbol{\mathcal{E}}_{m,t}$.
$\vartheta^{\mathrm{(void)}}_{m,i'',t}$ is the angle of the line perpendicular to the line connecting  $\boldsymbol{x}_{i'',t}$ and COM of the $m$-th void region at the time step $t$.
The schematic illustration of the definition $\vartheta^{(\mathrm{void})}_{m, i'', t}-\theta_{m, i'' ,t}$ is shown in Fig.~\ref{VoidSAC}(a).
Then, $\left< \bar{\beta} \right>$ is defined as the mean with respect to $m$, $t$, and the ten trials.
The orientational correlations $\left< \bar{\beta} \right>$ in the systems with rhomboidal particles are higher than those in the systems with elliptical particles as shown in Fig.~\ref{VoidSAC}(b).
The particles around void regions are more oriented along the void peripheries in the systems with rhomboidal particles than in the systems with elliptical particles.
The data for $a<3.0$ in the systems with rhomboidal particles and $a<4.0$ in the systems with elliptical particles could not be evaluated because the number of voids was too small.
We expect that the lower mobility due to almost close-packed structure in the systems with rhomboidal particles leads to the difference in the orientation around void regions.

Considering the alignment of the particles surrounding the void regions, we estimate void areas ${S}^{\mathrm{(void)}}_{\mathrm{(C)}}$ from $\lambda^{(\mathrm{void})}_{(\mathrm{C})}$ for each $a$. 
Figure~\ref{VoidSAC}(c) shows the schematic illustration for the definition of $S^{\mathrm{(void)}}_{\mathrm{(C)}}$.
Here, we expect that the angle difference between two particles with a distance of $\lambda^{(\mathrm{void})}_{(\mathrm{C})}$ along the void periphery is $\left( \arccos{\left( e^{-1} \right) } \right) /2$ considering the definition of the correlation length.
If the void regions have circular shapes, the void areas $S^{\mathrm{(void)}}_{\mathrm{(C)}}$ are estimated as
\begin{align}
\label{VoidSC}
S^{(\mathrm{void})}_{(\mathrm{C})}&=\pi \left( R^{(\mathrm{void})}_{(\mathrm{C})} \right) ^{2},
\end{align}
where
\begin{align}
\label{VoidRadius}
R^{(\mathrm{void})}_{(\mathrm{C})}=  \frac{ 2\pi \lambda^{(\mathrm{void})}_{(\mathrm{C})} }{ \arccos{\left( e^{-1} \right) }/2}.
\end{align}
Figure~\ref{VoidSAC}(d) shows ${S}^{\mathrm{(void)}}_{\mathrm{(C)}} / \left< \bar{S}^{\mathrm{(void)}} \right>$ against $a$ for the comparison between the observed mean void area and the estimated one.
The ratio for rhomboidal particles is almost constant, while that for elliptical particles greatly varies depending on $a$.
In addition, it should be mentioned that the systems with rhomboidal particles include more void regions with the winding number $+1$ than those with elliptical particles as shown in Fig.~\ref{VWR}.
These results suggest that the void areas in the systems with rhomboidal particles could be directly determined by the orientational correlation length.
Since the circularity of the void regions is important for the success in this estimation, the systematic analysis of the void shapes should be focused on in future work.

Our results show that not only the particle aspect ratio but also the particle shape change the number and size of void regions through direct collisions among particles.
In many studies on active nematics with hydrodynamic models, only the aspect ratio of the particles is included in their models~\cite{happel2012low}\cite{SK064}.
Our study clearly shows that both the details of the particle shapes and the aspect ratio of the particles influence the macroscopic collective motions, in a system of dry active nematics in which particles directly collide with each other.

Previously, Gro{\ss}man et al. reported the spatio-temporal pattern formation based on a similar model to the present one~\cite{SK067}.
They introduce a noise term in angular motion in their model, while the particles change their motion directions only by the collision with the other particles in our model.
We consider that we obtain the characteristic dimensionless parameter like Peclet number by varying the characteristic reversal time $\tau$ from the direct numerical simulation in future work.

\section{Conclusion}
\label{SecFive}
We simulated a model of active nematics, in which rhomboidal or elliptical particles interact by direct collision, and investigated by varying the particle aspect ratio to observe the correlation between the collective pattern and the particle shape.
The local density variance and the local nematic orientation increase with an increase in the aspect ratio both in the systems with rhomboidal particles and elliptical particles.
The void regions, where the particle number density is almost zero, appear only for the higher aspect ratio in both systems.
The systems with rhomboidal particles have the characteristic void size, which has a positive correlation with the aspect ratio, while the systems with elliptical particles do not have the clear characteristic void size and show a broad size distribution.
Most void regions are surrounded by particles whose long-axes are along with the peripheries of the void regions in the systems with rhomboidal particles.
Considering the proportionality between the observed void area and the void area estimated from the orientational correlation length, the size of the void regions might be determined by the particle shape and the aspect ratio through the difference in the orientational correlation length.

\begin{acknowledgments}
This work was supported by JST SPRING, Grant Number JPMJSP2109 (MS), by JSPS KAKENHI Grant Nos. JP21K13891 (HI), JP20H02712, JP21H00996, and JP21H01004 (HK), and by the Cooperative Research Program of ``NJRC Mater. \& Dev.'' No.20224003 (HK).
\end{acknowledgments}

\providecommand{\noopsort}[1]{}

\end{document}